\documentclass[final,5p,times,twocolumn]{elsarticle}

\usepackage{graphicx}
\usepackage{amssymb}
\usepackage{amsmath}
\usepackage{amsthm}
\usepackage[normalem]{ulem}
\biboptions{sort&compress}
\usepackage{slashed}
\usepackage{color,rotating}
\usepackage{bbm}
\usepackage{array}
\usepackage{multicol}
\usepackage[utf8]{inputenc}
\usepackage{subcaption}
\usepackage{multirow}
\usepackage{xcolor}

\allowdisplaybreaks

\begin{document}

\newcommand{\p}{\partial}
\newcommand{\hh}{{\widehat{h}}}
\newcommand{\bchi}{{\bar{\chi}}}
\newcommand{\btheta}{{\bar{\theta}}}
\newcommand{\ds}{{\slashed{\nabla}}}
\newcommand{\unit}{{\mathbbm{1}}}

\newcommand{\gb}{\bar{g}}
\newcommand{\Db}{\bar{D}}
\newcommand{\Rb}{\bar{R}}

\newcommand{\cO}{\mathcal{O}}
\newcommand{\cR}{\mathcal{R}}

\newcommand{\be}{\begin{equation}}
\newcommand{\ee}{\end{equation}}

\begin{frontmatter}

\title{The conformal sector of  Quantum Einstein Gravity beyond the local potential approximation}

\author{Alfio Bonanno}
\ead{alfio.bonanno@inaf.it}
\address{INAF, Osservatorio Astrofisico di Catania, via 
S.Sofia 78, I-95123 Catania, Italy ;\\
INFN, Sezione di Catania, via S. Sofia 64, I-95123, 
Catania, Italy.
}
\author{Maria Conti}
\ead{mconti@uninsubria.it}
\address{DISAT, Università degli Studi dell’Insubria, via 
Valleggio 11,  I-22100 Como, Italy ;
\\INFN, Sezione di Milano, 
Via Celoria 16, I-20133 Milano, Italy. }
\author{Dario Zappal\`a}
\ead{dario.zappala@ct.infn.it}
\address{INFN, Sezione di Catania, via S. Sofia 64, 
I-95123, 
Catania, Italy;\\
Centro Siciliano di Fisica Nucleare e Struttura della 
Materia, Catania, 
Italy.}

\begin{abstract}
The anomalous scaling of Newton's constant around the Reuter fixed  point is dynamically computed using the functional flow equation  approach. Specifically, we thoroughly analyze the flow of the most  general conformally reduced Einstein-Hilbert action. Our findings  reveal that, due to the distinctive nature of gravity, the  anomalous dimension  $\eta$ of the Newton's constant cannot be  constrained to have one single value: the ultraviolet  critical  manifold is characterized by a line of fixed points $(g_\ast(\eta), 
\lambda_\ast (\eta))$, with a discrete (infinite) set of  eigenoperators associated to each fixed point. More specifically, we find three ranges of $\eta$ corresponding to different  properties of both fixed points and eigenoperators and, in 
particular,  the range $ \eta < \eta_c \approx 0.96$  the ultraviolet critical manifolds has finite dimensionality. 
\end{abstract}


\end{frontmatter}

\section{Introduction}\label{sec:intro}
Understanding the structure of the ultraviolet (UV) critical manifold in Asymptotically Safe (AS) theories of gravity has been a formidable challenge despite substantial efforts. A striking departure from well-established examples, such as scalar theories in $d$ dimensions, lies in the unique role played by Newton's constant, $G_N$, in determining the scaling dimension of the gravitational field in the action:

\begin{equation}
\label{action1}
S[g_{\mu\nu}] = -\frac{1}{16 \pi} \int d^4 x \frac{1}{G_N} \sqrt{g} R
\end{equation}

In this context, the inverse of $G_N$, equivalent to the wavefunction renormalization function $Z$ for the graviton, assumes a central role. Percacci and Perini highlighted a crucial distinction from the scalar field case, where, despite its technical inessentiality \cite{Percacci:2004sb}, $1/G_N$ should scale as the inverse of the cutoff square. Consequently, the anomalous dimension of the field obeys the equation:
\begin{equation}
\label{classical}
\eta = \beta_N - 2.
\end{equation}
Here, $\beta_N$ denotes the beta-function for Newton's constant which can be obtained from (\ref{action1}).
At the fixed point (FP), where $\beta_N$ vanishes, the anomalous dimension $\eta$ takes on the value of -2 \cite{Reuter:1996cp, Percacci:2017fkn, Reuter:2019byg}. The value $\eta = -2$ merely reflects the classical (negative) natural dimension of the background Newton's constant. In contrast, the anomalous dimension associated with graviton fluctuations, denoted as $\eta_h$, emerges dynamically and takes a value close to one ($\eta_h \approx 1.02$ as reported in \cite{Bonanno:2021squ}).

The central question we aim to tackle in this paper is whether we can transcend the definition provided in equation  (\ref{classical}), which is based on  a $\beta$-function approach, (i.e. implying polynomial truncation of  the theory space), and propose a dynamic determination of the anomalous dimension of the background field, similar to the case of inessential couplings in standard quantum field theory (QFT). To overcome the limitations of equation (\ref{classical}), it is evident that we must approach the problem at a non-perturbative level, taking into account all possible operators compatible with the general group of diffeomorphisms.

As originally pioneered by Wilson and Wegner \cite{Wilson:1971bg, Wegner:1972ih}, the Renormalization Group (RG), when implemented with the aid of Functional Flow Equations, serves as a powerful tool for exploring the structure of the UV critical manifold at a non-perturbative level \cite{Nicoll:1977hi, Polchinski:1983gv, Wetterich:1992yh, Morris:1993qb}.
At the heart of this approach lies the Effective Average Action $\Gamma_k$, representing the effective action of the system at the scale $k$, which converges to the familiar Effective Action in the limit as $k\rightarrow 0$.

In recent years, the Renormalization Group, implemented with Functional Flow Equations, has proven invaluable for exploring the properties of the UV critical manifold in Asymptotically Safe (AS) theories of gravity beyond the framework defined in (\ref{action1}). Researchers have turned to theories of the form:
\begin{equation}
\label{fr}
\Gamma_k[g_{\mu\nu}] = \frac{1}{16\pi}\int d^4 x\sqrt{g} f_k(R),
\end{equation}
aiming to move beyond polynomial truncation. For an extensive review of this approach, please refer to \cite{Morris:2022btf}. The rationale behind (\ref{fr}) draws parallels with the Local Potential Approximation (LPA) in scalar theories, where exact flow equations for the potential $V(\phi)$ have been derived and successfully solved, leading to highly accurate studies of critical properties in dimensions $d=3$ and $d=4$ . Despite substantial computational efforts dedicated to analyzing scaling solutions (i.e., fixed point)  of the flow equation for (\ref{fr}), progress in moving beyond Eq.  (\ref{classical}) has remained somewhat limited. This challenge  is further compounded by the issue of background independence.

The Wetterich equation is grounded in the Effective Average Action  $\Gamma_k$, which, in the context of the background field method,  depends on two distinct metrics: the full metric $g_{\mu\nu}$ and the background field metric $\bar{g}_{\mu\nu}$. While in standard gauge theory all correlation functions of the fundamental field can be derived from ``on shell" background field correlation functions with the aid of a gauge-fixing term invariant under a background transformation, in gravity, particularly following the tenets of loop quantum gravity \cite{Ashtekar:2004eh}, background independence carries a more profound significance. Here, full background independence entails that no metric should hold any privileged status at any stage of the calculation.

Given the formidable technical complexity of achieving full background independence, advancement in this direction necessitates significant simplifications. For instance, building upon the approach initiated by \cite{Reuter:2008wj,Reuter:2008qx}, several studies have explored a conformally reduced version of the 
theory  \cite{Machado:2009ph,Bonanno:2012dg}, demonstrating that the conformal factor plays a pivotal role in generating the Reuter fixed point with a finite number of relevant directions, even 
beyond the standard Einstein-Hilbert truncation  \cite{Bonanno:2023ghc}.

In contrast, Morris and his collaborators have addressed the more  rigorous requirement of full background independence directly within the flow equation framework in a series of works \cite{Dietz:2015owa,Labus:2016lkh,Dietz:2016gzg}. Particularly, in  \cite{Dietz:2016gzg}, the use of a specific cutoff function  allowed for the examination of the ultraviolet critical manifold  of the conformally reduced theory beyond the LPA approximation  employed in \cite{Reuter:2008qx}. A significant finding from this  work is the identification of a line of fixed points,  {each with a  continuous set of relevant eigendirections}. Taken at face value,  this result suggests that the theory is renormalizable but not predictive, given the infinite number of relevant directions.

In this paper, we pursue an alternative path. We employ a spectrally-adjusted cutoff  \cite{Bonanno:2022edf} previously utilized in scalar theory and quantum gravity investigations
\cite{Bonanno:2000yp,Bonanno:2004sy,Bonanno:2012dg}, to surpass the LPA approximation presented in \cite{Reuter:2008qx}. Consequently, we adopt a single metric approximation for the renormalized flow.  By virtue of our cutoff selection, we are able to obtain analytical solutions. While we confirm the presence of a line of fixed points,  we observe a discrete spectrum of eigendirections. Notably, the UV critical manifold exhibits finite dimensionality for $ \eta < \eta_c \approx 0.96$.

The structure of the paper is the following. In Sec. 2 the general framework and the approximation scheme are  discussed; in Sec. 3 the numerical analysis employed for the  determination of the fixed points and for the  eingenvalue spectra is presented; the conclusions are  reported in Sec. 4.

\section{General setup and differential equations in the flat projection}
We start by considering the well known Euclidean Einstein-Hilbert 
action \cite{Bonanno:2012dg}
\small
\begin{equation}
S^{EH}[g_{\mu\nu}]=-\frac{1}{16\pi}\int d^d x\sqrt{g}\, G^{-1}\,(R-2\Lambda),
\label{creh}
\end{equation}
\normalsize
which, by Weyl rescaling $g_{\mu\nu}=\phi^{2\nu}\hat{g}_{\mu\nu}$ where $\nu$ is a parameter taken as $\nu=\frac{2}{d-2}$, can be written as
\small
\begin{align}
S^{EH}_k[\phi]&=\int d^dx\, \frac{\sqrt{\hat{g}} \, Z_k}
{2}\,\left(\hat{g}^{\mu\nu}\,\partial_\mu\phi\partial_\nu\phi +A\,\hat{R}\,\phi^2-
4A\,\Lambda_k\phi^{\frac{2d}{d-2}}\right),\label{SEHweyl}
\end{align}
\normalsize
where $\hat{R}\equiv R(\hat{g})$, $A=A(d)=\frac{d-2}{8(d-1)}$ and
\small
\begin{equation}
Z_k=-\frac{1}{2\pi \, G_k}\, \frac{d-1}{d-2}
\label{ZG}
\end{equation}
\normalsize
and it is assumed that $\Lambda_k$ and $G_k$ (and $Z_k$ through Eq. (\ref{ZG}))
are RG scale $k$-dependent parameters. It is then convenient
to consider a general action of the type
\small
\begin{equation}
S_k[\phi]=\int d^d x \sqrt{\hat{g}}\,\left(\frac{1}{2}Z_k\; 
\hat{g}^{\mu\nu}\partial_\mu\phi\,\partial_\nu\phi+V_k[\phi]\right),
\label{Sgen}
\end{equation}
\normalsize
where $V_k[\phi]\equiv Z_kU_k[\phi]$, with $U[\phi]$ a potential term for  the action. Then,  the parallel between Eqs.
(\ref{SEHweyl}) and (\ref{Sgen}) is evident.\\ 
Following \cite{Bonanno:2000yp,Bonanno:2012dg} the proper-time flow equation  reads
\begin{equation}
\partial_t\, {\Gamma}_k [f; \chi_B] = - \frac{1}{2}\,  {\rm Tr} \int_0^{\infty} \frac{ds}{s}\, \partial_t\, \rho_{k}\, 
\exp\left({-s \frac{\delta^2 \Gamma_k [f; \chi_B]}{\delta {f}^2} }\right),
\label{floweq}
\end{equation}
where $t\equiv\log k$ is the RG time and the original field $\phi$ is split into a constant background component $\chi_B$ and a dynamical fluctuation $\bar{f}$ : ${\phi(x)=\chi_B+\bar{f}(x)}$.  Moreover, a reference metric $\hat{g}_{\mu\nu}$ is chosen through a Weyl rescaling $\bar{g}_{\mu\nu}=\chi_B^{2\nu}\hat{g}_{\mu\nu}$, with constant $\chi_B$ and $\nu=2/(d-2)$, where $d$ is the space-time dimension and, because of this definition, the two momenta scales (defined as the  Eigenvalues of the $-\hat{\Box}$ and $-\bar{\Box}$ respectively) are usually related through
\begin{equation}
\hat{k}^2=\chi_B^{2\nu}\bar{k}^2.\label{khatkbar}
\end{equation}
and the conformal factor is considered as dimensionless.
However, as discussed in the introduction, in our investigation we consider an anomalous scaling for the $\chi_B$ and we write
\begin{equation}
\chi_B =\psi_B \, k ^{\frac{\eta}{2}}.\label{chipsi}
\end{equation}
being $\psi_B$  a dimensionless field and $\eta$ the  anomalous dimension. This choice entails some differences with respect to \eqref{khatkbar}. While the identity ${\hat{\Box}=\chi_B^{2\nu}\,\bar{\Box}}$ always holds true, in order for $\hat{\Box}$ to show the correct dimensionality of $k^2$ $({[\hat{\Box}]=k^2)}$, it must imply: ${[\bar{\Box}]=k^{2-\eta\nu}}$. Hence, the  relation between momenta built with the reference metric and the background metric 
must be of the following form:
\begin{equation}
\hat{k}^2=\chi_B^{2\nu}\bar{k}^{2-\eta\nu}, \label{etaminustwo}
\end{equation}
which for ${d=4}$, i.e. ${\nu=1}$, simply reduces to 
${\hat{k}^2=\chi_B^2\bar{k}^{2-\eta}}$.\\
The smooth infrared (IR) regulator $\rho_k=\rho_k[\chi_B]$ 
(repeatedly analyzed in  the literature \cite{
Bonanno:2000yp,Zappala:2001nv,Bonanno:2004sy,Litim:2010tt,
Bonanno:2012dg,Bonanno:2022edf}),
which depends  on the background field only $\chi_B$,  and whose role is to  restrict the integration to the modes with 
$p\geq k$, is taken  as in \cite{Bonanno:2012dg}: because of the new relation \eqref{etaminustwo}, it now assumes the form
\begin{align}
&\partial_t\, \rho_k(s,n)=-\frac{2}{\Gamma(n)}(s\, n\, Z_k\,k^{2-\eta\nu}\, \chi_B^{2\nu})^n e^{-s\, 
n\,Z_k\, k^{2-\eta\nu} \chi_B^{2\nu}},\label{dtregformnew}
\end{align}
where $n$ is a positive integer number controlling the sharpness of the  cutoff and $\Gamma(n)$ is the Gamma function. Following the background field method, the trace must always be 
performed on the modes constructed with the background metric  $\bar{g}_{\mu\nu}$, i.e. the trace must be taken on the spectrum of  $-\bar{\Box}$ \cite{Reuter:2008wj,Reuter:2008qx}.
The spectral adjustment represented by the presence of the  renormalization function $Z_k$ inside the exponential allows to correctly perform a cut on the eigenvalues of the operator $-\bar{\Box}$ when the original theory contains corrected operators of the type $-Z_k\,\hat{\Box}$.

The implementation in Eq. \eqref{Sgen} of an  expansion in powers of the field fluctuation \begin{equation}
\bar{f}=\phi(x)-\chi_B\label{barf}
\end{equation} 
around the constant background $\chi_B$, leads to the following form for the left hand side of the flow equation:
\begin{equation}
\partial_t \Gamma_k=\int d^d x \sqrt{\hat{g}}\left(\frac{1}{2}(k\partial_k Z_k)[\chi_B] 
\bar{f}\hat{\Box}\bar{f}+k\partial_k V_k[\chi_B]  \right).\label{lhs}
\end{equation}
The right hand side of the flow equation can instead be expressed as
\begin{align}
-\frac{1}{2}\int \frac{ds}{s}\int d^d x \sqrt{\bar{g}}\int \frac{d^d \bar{p}}
{(2\pi)^d}(k\partial_k \rho_k)\langle x|\bar{p}\rangle\langle \bar{p}|e^{-
s\Gamma_k^{(2)}}|x\rangle,\label{rhsexpand}
\end{align}
where the trace is performed on the modes of $-\bar{\Box}$. This expression can be expanded through Baker-
Campbell-Hausdorff expansion up to quadratic terms in $\bar{f}$, leading to
\small
\begin{align}
&-\frac{1}{2}\int \frac{ds}{s}\int d^dx\,\chi_B^{d\nu}\sqrt{\hat{g}}\,(k\partial_k 
\rho_k) \int\frac{d^d\bar{p}}{(2\pi)^d}e^{-s(Z_k\chi_B^{2\nu}\bar{p}^{2-
\eta\nu}+V_k^{(2)})}\times\nonumber\\
&\quad\quad\times\left(1-sB_1+\frac{s^2}{2!}B_2-\frac{s^3}{3!}B_3+\frac{s^4}
{4!}B_4\right)\label{BCH}
\end{align}
\normalsize 
where the identity $\sqrt{\bar{g}}=\chi_B^{d\nu}\sqrt{\hat{g}}$ was inserted. The 
expressions denoted with $B_i$ with $i=1,2,3,4$ explicitly read
\small
\begin{align}
B_1&=\left[Z'[\chi_B]\hat{p}^2\bar{f}+\frac{Z''[\chi_B]}{2}\hat{p}^2\bar{f}^2-
\frac{Z''[\chi_B]}{2}\bar{f}\hat{\Box}\bar{f}+\right.\nonumber\\
&\quad\left.\left.+\bar{f}V^{(3)}[\chi_B]+\frac{\bar{f}^2}{2}V^{(4)}
[\chi_B])\right]\right\rvert_{\hat{\Box}=\chi_B^{2\nu}\bar{\Box},\,\,\, 
\hat{p}^2=\chi_B^{2\nu}\bar{p}^{2-\eta\nu}};\label{B1}
\end{align}
\normalsize 
\\
\small
\begin{align}
B_2&=\bigg[(Z'[\chi_B])^2\hat{p}^4\bar{f}^2-(Z'[\chi_B])^2\left(2+\frac{1}
{d}\right)\hat{p}^2\bar{f}\hat{\Box}\bar{f}+\nonumber\\
&\quad+2Z'[\chi_B]V^{(3)}[\chi_B]\hat{p}^2\bar{f}^2-2Z'[\chi_B]V^{(3)}
[\chi_B]\bar{f}\hat{\Box}\bar{f}+\nonumber\\
&\quad\left.+\bar{f}^2(V^{(3)}
[\chi_B])^2\bigg]\right\rvert_{\hat{\Box}=\chi_B^{2\nu}\bar{\Box},\,\,\, 
\hat{p}^2=\chi_B^{2\nu}\bar{p}^{2-\eta\nu}};\label{B2}
\end{align}
\normalsize 
\\
\small
\begin{align}
B_3&=\left[-Z[\chi_B](Z'[\chi_B])^2\left(1+\frac{4}
{d}\right)\hat{p}^4\bar{f}\hat{\Box}\bar{f}+\right.\nonumber\\
&\quad-Z[\chi_B]Z'[\chi_B]V^{(3)}[\chi_B]\left(2+\frac{4}
{d}\right)\hat{p}^2\bar{f}\hat{\Box}\bar{f}+\nonumber\\
&\quad\left. -Z[\chi_B](V^{(3)}
[\chi_B])^2\bar{f}\hat{\Box}\bar{f}\bigg]\right\rvert_{\hat{\Box}=\chi_B^{2\nu}\bar
{\Box},\,\,\, \hat{p}^{2}=\chi_B^{2\nu}\bar{p}^{2-\eta\nu}};\label{B3}
\end{align}
\normalsize 
\\
\small
\begin{align}
B_4&=\left[-(Z[\chi_B])^2(Z'[\chi_B])^2\frac{4}
{d}\hat{p}^6\bar{f}\hat{\Box}\bar{f}+\right.\nonumber\\
&\quad-(Z[\chi_B])^2Z'[\chi_B]V^{(3)}[\chi_B]\frac{8}
{d}\hat{p}^4\bar{f}\hat{\Box}\bar{f}+\nonumber\\
&\quad-\left.\left.(Z[\chi_B])^2(V^{(3)}[\chi_B])^2\frac{4}
{d}\hat{p}^2\bar{f}\hat{\Box}\bar{f}\right]\right\rvert_{\hat{\Box}=\chi_B^{2\nu}\
\bar{
\Box},\,\,\, \hat{p}^2=\chi_B^{2\nu}\bar{p}^{2-\eta\nu}}.\label{B4}
\end{align}
\normalsize 
As already discussed, in Eqs. \eqref{B1}-\eqref{B4} the two relations ${\hat{\Box}=\chi_B^{2\nu}\bar{\Box}}$ and ${\hat{p}^2=\chi_B^{2\nu}\bar{p}^{2-\eta\nu}}$ 
are enforced.  The latter was also applied to the momenta appearing in Eq. (\ref{dtregformnew}),  which specifies the form of the regulator.  After integrating over the background momenta $\bar{p}$ and over the variable $s$ in \eqref{BCH}, a comparison with the left hand side of the flow equation in \eqref{lhs} allows to identify the flow  equations for $V_k$ and $Z_k$. Apart from a common coefficient $C$
\small
\begin{equation}
C=\frac{2^{-d} \pi ^{-d/2} \Gamma \left(\frac{d-\eta  \nu +2}{2-\eta  \nu }\right) n^{\frac{d}{2-\eta  \nu
   }} \Gamma \left(\frac{d}{\eta  \nu -2}+n\right)}{\Gamma \left(\frac{d}{2}+1\right) \Gamma (n)},
\end{equation}
\normalsize
the dimensional equations (with general dimension $d$ and shaping parameter $n$) for $V_k$ and $Z_k$ are:
\begin{equation}
\partial_t V_k=n^{\frac{d}{f_0}+n} \psi ^{d \nu } k^{-f_0 n} W_1^{-\frac{d}{f_0}-n} Z_k^{\frac{d}{f_0}+n} \chi_B ^{2 \nu 
   \left(\frac{d}{f_0}+n\right)}\label{delirioV}
\end{equation}
and
\footnotesize
\begin{align}
&\partial_t Z_k= -\frac{1}{3 f_0^4} q_0\, n^{\frac{d}{f_0}+n}\, Z_k^{\frac{d}{f_0}+n-1} k^{\frac{\eta  \nu  (d+f_0)}{f_0}+2 n} \chi_B ^{\frac{2 \nu  (d+n f_0-4)}{f}}\psi ^{d \nu }\, W_2^{-\frac{d+f_0 (n+3)}{f_0}}\times\nonumber\\
&\times\,\Biggr( \chi_B ^{\frac{8 \nu }{f_0}} k^{2 \eta  \nu } \left({V''_k}^2 \left(3 f_0^3 Z_k Z''_k+u_0
   {Z'_k}^2\right)+r_0 V^{(3)}_k Z ((f_0+1) s_0 V^{(3)}_k Z_k-2 t_0 V''_k Z'_k)\right)+\nonumber\\
&\quad-2 n\,Z_k\, k^{l_0}\, \chi_B ^{\frac{2 l_0 \nu }{f_0}} \left(r_0\, t_0\, V^{(3)}_k Z_k Z'_k-V''_k \left(3 f_0^3 Z_k Z''_k+u_0\,
   {Z'_k}^2\right)\right)+\nonumber\\
&\quad+k^4\,n^2\,Z_k^2\, \chi_B ^{\frac{4 \eta  \nu ^2}{f_0}} \left(3 f_0^3 Z_k Z''_k+u_0 {Z'_k}^2\right)\Biggr)\, , \label{delirioZ}
\end{align}
\normalsize
where the derivatives are taken with respect to the field $\phi$. Inside \eqref{delirioV} and \eqref{delirioZ}, the following quantities have been introduced for brevity:
\small
\begin{align}
&W_1=n\, Z_k\, \chi_B ^{2 \nu }\, k^{2-\eta  \nu }+V''_k\\
&W_2=k^2\, n\, Z_k\, \chi _B^{2 \nu }+k^{\eta  \nu } V''_k.
\end{align}
\normalsize
Together with $W_1$ and $W_2$, the functions $f_0$, $g_0$, $h_0$ 
and $l_0$ have been defined as 
$f_0=\eta  \nu -2$, $g_0=3 \eta  \nu 
-5$, $h_0=5 \eta  \nu -9$ and $l_0=2+\eta\nu$, 
that appear in the following definitions 
\small
\begin{align}
q_0 &= d + f_0 n\\
r_0 &= d + f_0 (1 + n)\\
s_0 &= d + f_0 (2 + n)\\
t_0 &= d (1 + f_0) + f_0 g_0\\
u_0 &=d^2 (f_0+1)+d f_0 h_0+f_0^2
\end{align}
\normalsize
and the `$0$' subscripts have been introduced 
to avoid confusion with other variables. 
We now introduce $Y_k$, $z_k$, $X$ as the 
dimensionless counterparts of 
the potential $V_k$, the renormalization function $Z_k$ and the field $\phi$ respectively as follows:
\begin{align}
& V_k[\phi]=Y_k[X]k^d\label{VadimY} \\
& Z_k[\phi]=z_k[X]k^{d-2-\eta}\label{etareutereq}\\
& \phi= X\,k^{\frac{\eta}{2}}\label{phiadimX},
\end{align}
where $\eta$ represents the anomalous dimension of \textit{both} 
fields $\chi_B$ and $\phi$, as shown by equations \eqref{chipsi} 
and \eqref{phiadimX} respectively.

Following the single field approximation \cite{Bonanno:2012dg}, 
once the fluctuation $\bar f$ defined in \eqref{barf} is 
integrated out, the  expectation value of the dimensionless 
field $X$, introduced in Eq. \eqref{phiadimX}, 
is identified with the the background $\psi_B$, 
which, for the sake of simplicity, from now on is 
indicated as $x$:
\begin{equation}
X=\psi_B\equiv x\,  .
\end{equation}

In terms of the dimensionless quantities $Y_k$ and $z_k$ in $d=4$, 
equations \eqref{delirioV} and \eqref{delirioZ} now read more 
easily as:
\small 
\begin{equation}
\partial_t Y_k(x)=-4 Y_k + \frac{\eta\, x\, Y'_k}{2} +n^{\frac{4}{\eta -2}+n} 
x^{\frac{10 \eta }
{\eta -2}+2 n} z_k^{\frac{4}{\eta -2}+n}\, Q_1^{-\frac{4}{\eta -2}-n-3}\, Q_2^3
\label{partY}
\end{equation}
\normalsize

\begin{figure*}
\centering
\includegraphics[scale=0.5]{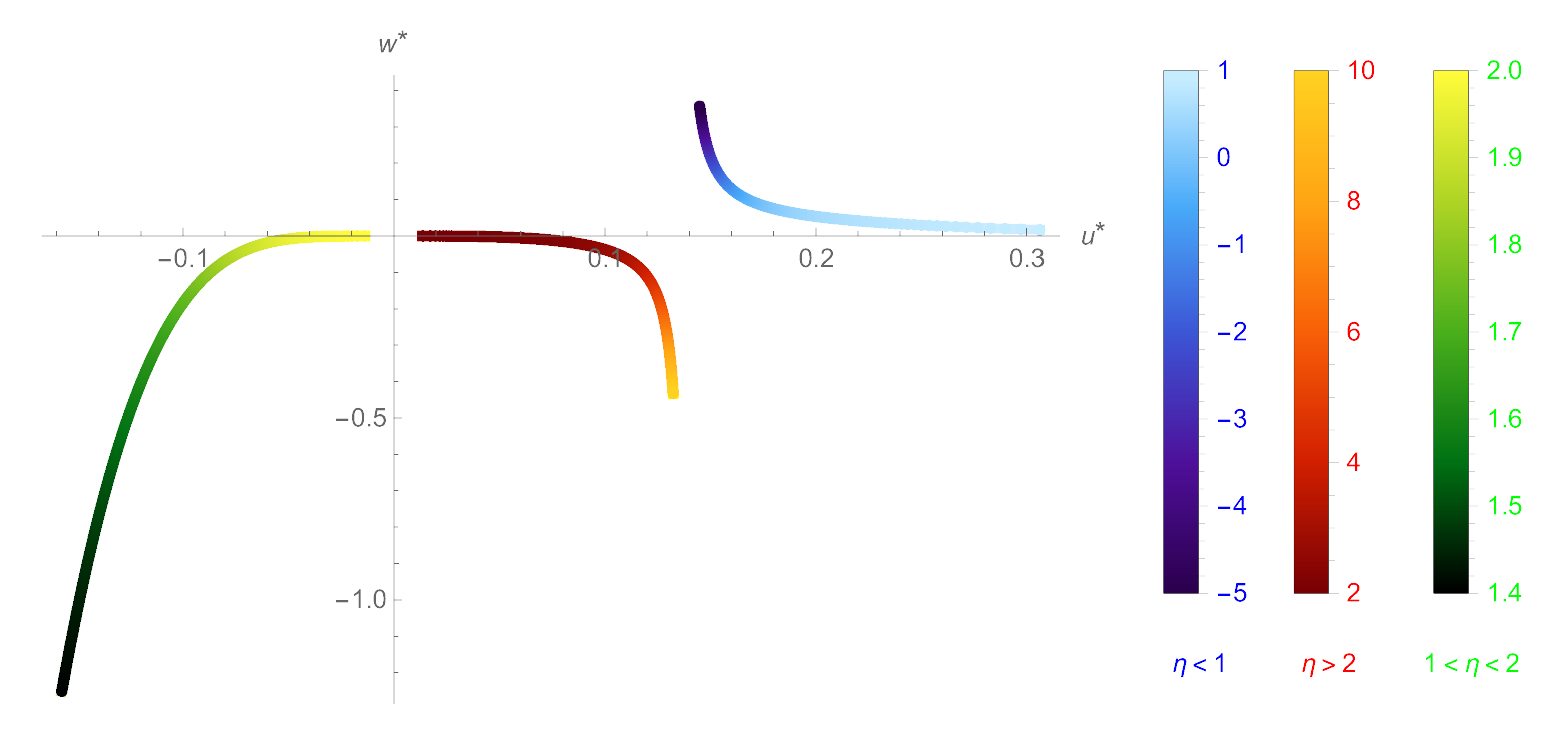}
\caption{Fixed point line for different values of the anomalous dimension $\eta$.\\
Blue curve: $\eta<1$; red curve: $\eta>2$; green curve: $1<\eta<2$.}
\label{fig:line}
\end{figure*}

\noindent 
and
\small
\begin{align}
&\partial_t z_k(x)=
(\eta -2)\, z_k+\frac{\eta  x z'_k}{2} -\frac{n^{\frac{4}{\eta -2}+n} ((\eta -2) n+4)}
{3 (\eta -2)^4}
x ^{\frac{4 \eta }{\eta -2}+2 n}
\times\nonumber\\
&\times  z_k^{\frac{4}{\eta -2}+n-1} \,
Q_1^{-\frac{4}{\eta -2}-n-3} \, \Bigg\{
(\eta -1) (Y^{(3)}_k)^2 z_k^2 
\Big(   (\eta -2) (n+1)+4   \Big) 
\times\nonumber\\
&\times
\Big( (\eta -2)(n+2)  +4  \Big)
-\Big(  4 (\eta -1)  +   (\eta -2)(3 \eta -5)  \Big) 
\times\nonumber\\
&\times
\Big(  (\eta -2)(n+1)+4  \Big)
 2\,Y^{(3)}_k z_k z'_k \, Q_1
+  \Big [ 3 (\eta -2)^3 z_k z''_k +
\nonumber\\
&+\Big(  (\eta -2)^2+4 (5 \eta -9) (\eta -2)+16 (\eta -1) \Big) (z'_k)^2\Big]
\, Q_1^2 \;
\Bigg\}
\label{partz}
\end{align}
\normalsize
where 
\begin{equation}
Q_1=\left(n x ^2 z_k+Y''_k\right)
\end{equation}
and 
\begin{equation}
Q_2=\left(n x ^{-\frac{4}{\eta 
-2}} z_k+x ^{-\frac{2 \eta }{\eta -2}} Y''_k\right)
\end{equation}
and the  derivatives appearing in the right hand side of \eqref{partY} and \eqref{partz} 
are taken with respect to the dimensionless field $X$.\\
Eqs. \eqref{partY} and \eqref{partz} are lengthy and rather involved 
as far as their numerical treatment is concerned, 
especially in the case of large $n$. 
Therefore we shall take the limit  $n\to +\infty$ in the two flow equations,
which are turned into a  more compact form, as all the troublesome powers 
of $n$  in \eqref{partY} and \eqref{partz} are replaced in this limit by 
$n$-independent exponential factors that are easier to handle numerically. 
In this limit, the two flow equations read:
\begin{equation}
\partial_t Y_k(x)=-4 Y_k + \frac{1}{2} \eta  x Y'_k+x ^4 e^{-\frac{Y''_k}{x ^2 
z_k}}\label{eqfinY}
\end{equation}
and
\begin{align}
&\partial_t z_k(x)=(\eta -2) z_k+\frac{1}{2} \eta  x z'_k-\frac{e^{-\frac{Y''_k}{x 
^2 z_k}}}{3 (\eta -2)^3 x ^2 z_k^2} \times\nonumber\\
& \times\Big[  (\eta -2)^2 (\eta -1) (Y^{(3)}_k)^2 +
\nonumber\\
& +2 \left(-3 \eta ^3+13 \eta ^2-20 \eta +12\right) x ^2Y^{(3)}_k z'_k+ 
\nonumber\\
& +x ^4 
\left(3 (\eta -2)^3 z_k z''_k+\left(21 \eta ^2-64 \eta +60\right) (z'_k)^2\right)\Big]
\label{eqfinz}
\end{align}
In the next  Section we shall analyse the fixed point structure of these flow  equations including the central role of the anomalous  dimension $\eta$ and then, the nature of the  related  eigendirections.


\section{The role of the anomalous dimension}\label{sec:results}
\subsection{The fixed point line structure}\label{sec:FPline} In 
this Section we focus on the fixed point solutions of  the flow 
equations \eqref{eqfinY} and \eqref{eqfinz}, i.e.  $\partial 
Y_k(x)= \partial z_k(x)= 0$,
for the  generalization of the conformally reduced Einstein-
Hilbert (CREH)  truncation shown in \eqref{SEHweyl},
with $d=4$ and with 
flat  curvature $\hat{R}=0$.
It is easy to check that the functions
\small
\begin{align}
\label{numsol1}
&z=-\frac{1}{w} \\
&Y=\frac{u}{w} x^4
\label{numsol2}
\end{align}
\normalsize
(where $w$ and $u$ are numerical coefficients), 
are such that all dependence on $x$ gets canceled both in Eqs. 
\eqref{eqfinY} and \eqref{eqfinz}, so that the two differential 
equations  reduce to simple numerical equations. In addition, 
from \eqref{creh}, \eqref{SEHweyl}, \eqref{ZG} and 
\eqref{numsol1}, \eqref{numsol2}, 
one easily deduce that $u$ and $w$ correspond respectively 
to the dimensionless 
cosmological constant $\lambda$ and Newton's constant $g$.

It is crucial to remark that all numerical searches performed
to determine alternative solutions $Y(x)$ and $z(x)$ did not 
produce any positive result, leaving the monomial functions
in  \eqref{numsol1} and \eqref{numsol2} as the only acceptable
fixed point solution among the entire set of real,
polynomial or non-polynomial, functions.

After replacing the functions \eqref{numsol1}, \eqref{numsol2} 
in  Eqs. \eqref{eqfinY} and \eqref{eqfinz}, we get the following  
equations in terms of the three parameters $u$, $w$, and $\eta$ :
\small
\begin{align}
\label{algsol1}
&e^{12\,u}+\frac{2u\,(-2+\eta)}{w}=0\\
&\frac{2-\eta}{w} -\frac{192\,e^{12\,u}\,u^2\,(-1+\eta)}{-2+\eta}=0
\label{algsol2}
\end{align}
\normalsize

\begin{figure*}
\centering
\includegraphics[scale=0.5]{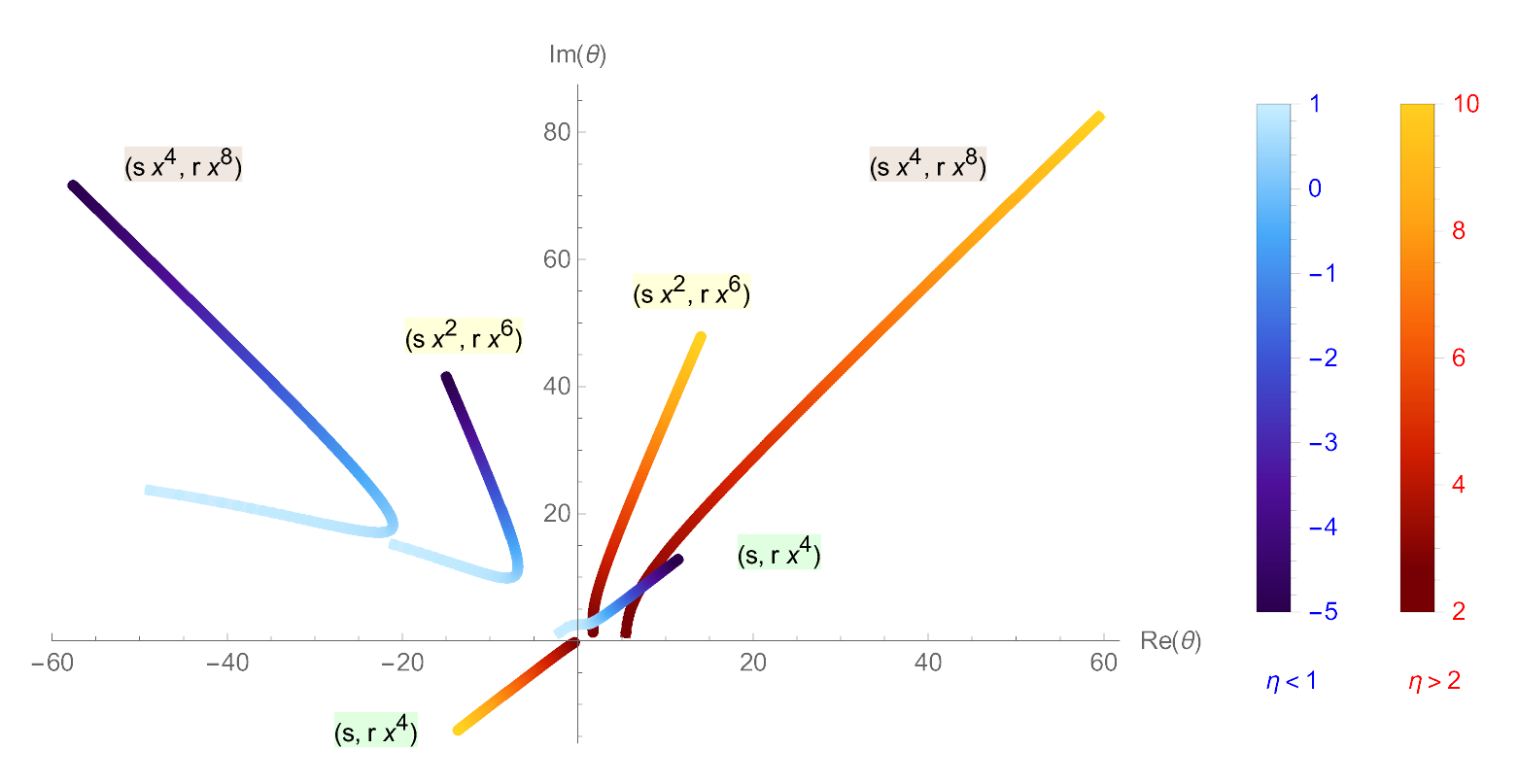}
\caption{Eigenvalues for three different types of eigenfunctions: 
${(f,h)=\{(s,r\,x^4),
(s\,x^2, r\,x^6),(s\,x^4, r\,x^8)\}}$, for different values of 
the anomalous dimension $\eta$.
Blue curve: ${\eta<1}$; red curve: ${\eta>2}$. As eigenvalues 
always show the form 
${\theta_{1,2}=\text{Re}(\theta)\pm i\, \text{Im}(\theta)}$ for 
these values of 
$\eta$, only one of the two Eigenvalues is plotted.}
\label{fig:mix1}
\end{figure*}

Clearly, one of the three parameters cannot be determined 
and we choose to parameterize the solutions of Eqs. \eqref{algsol1}, 
\eqref{algsol2}, in terms of the anomalous dimension
$\eta$.
The corresponding solutions $u^*(\eta)$ and $w^*(\eta)$ are 
reported in Fig. \ref{fig:line}. It is evident from Eqs. \eqref{algsol1}, 
\eqref{algsol2},
that   $\eta=2$ is a singular point, whose origin can be 
traced back to the dimensions of the background field
and to the exponent of the momentum $\bar k$ in 
Eq. \eqref{etaminustwo}. However, as can  be checked in
Eqs. \eqref{algsol1} and \eqref{algsol2}, $u^*$ and $w^*$
are both regular and vanish in the limit $\eta\to 2$
(In Fig. \ref{fig:line}, $u^*$ and $w^*$ are plotted up to
$\eta=2\pm 0.001$). From Eqs. \eqref{algsol1}, 
\eqref{algsol2}, we also notice that, in the limit 
$\eta\to 1$, at least one of the two solutions
$u^*$ and $w^*$  diverge,
thus making the point $\eta=1$ singular; in addition,
when $\eta\to \pm \infty$, the parameter $w^*$ diverges.

Therefore, Fig. \ref{fig:line} shows  three disconnected lines of 
fixed points parameterized in terms of the continuous variable 
$\eta$, with two singular points at $\eta=1$ and $\eta=2$. This 
is quite different from the picture  of a simple scalar quantum 
field theory where, in the same kind of analysis, one can
constrain $w^*=1$, because any rescaling of  $w^*$ would imply 
a physically irrelevant rescaling of $u^*$ and of the field, with 
no change in the spectrum of the eigenvalues of the linearized 
flow equations around the fixed point solution. In this case, the 
inclusion of the same anomalous dimension for \textsl{both} 
fields $\phi$ and $\chi_B$, together with the identification of 
$u^*$ and $w^*$ with the physically meaningful quantities 
$\lambda^*$ and $g^*$ makes the rescaling of $w^*$ an appreciable 
operation. We conclude that each
FP of the lines in Fig. \ref{fig:line}, 
corresponds to a distinct solution.

From Fig. \ref{fig:line}, we notice that the solutions
$u^*$  and $w^*$ maintain a definite sign within each of the 
three ranges $\eta<1$, $1<\eta<2$ and $2<\eta$, but they can
switch sign when passing from one range to another.
Finally,
we recall that the point  $\eta=0$ corresponds to the usual 
formalism of literature \cite{Bonanno:2012dg}, as 
\eqref{VadimY} and \eqref{etareutereq} reproduce the known 
scaling of $V_k$ and $Z_k$. In other words, $\eta=0$ must 
correctly reproduce the  Reuter FP as in \cite{Bonanno:2012dg} , characterized by $g^*>0$ 
and $\lambda^*>0$ and, in fact, within our computation, the 
Reuter FP  with $\eta=0$ is located at $w^*=0.086$ 
and $u^*=0.173$.

\begin{figure*}
\centering
\includegraphics[scale=0.5]{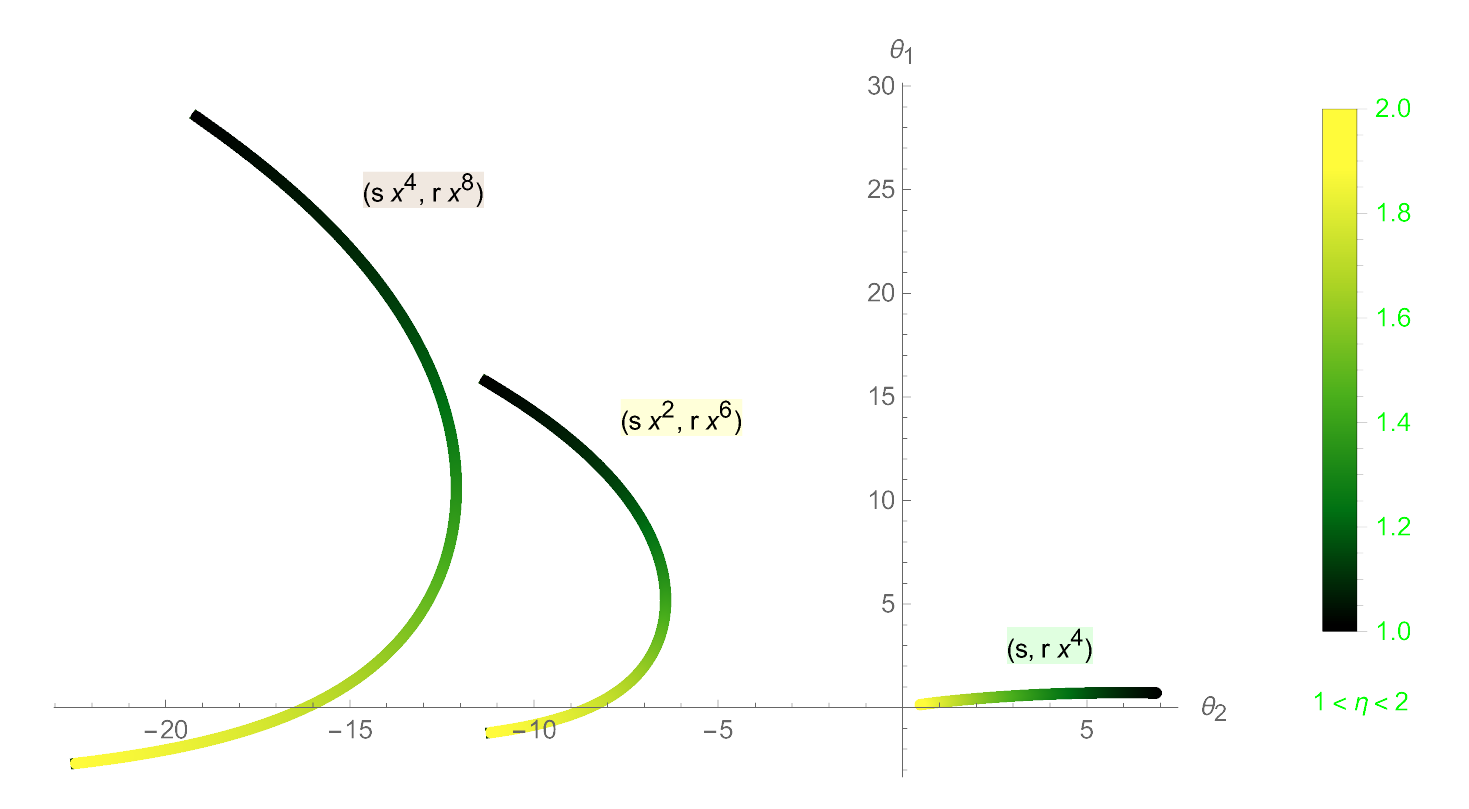}
\caption{Eigenvalues for three different types of Eigenfunctions: 
${(f,h)=\{(s, r\,x^4),(s\,x^2, r\,x^6),(s\,x^4, r\,x^8)\}}$, for different values 
of  the anomalous dimension ${1<\eta<2}$.  
The two eigenvalues associated to each value of $\eta$ 
within this range, are real and their values are 
respectively reported on the two axes of the plot.}
\label{fig:mix2}
\end{figure*}

Although it seems natural to expect $u^*,w^*>0$, 
which in our computation is recovered only when $\eta<1$, 
we do not have any 
argument to reject the negative solutions as long as they 
correspond  to UV fixed points whose outgoing trajectories 
lead in the IR limit to physically acceptable positive 
values of  $\lambda$ and $g$. Therefore the following 
mandatory step is the study of the eigenfunction spectrum 
of our flow equations.

\subsection{Eigenfunction spectrum}\label{sec:Eigenfuncs}
The set of eigenfunctions,
that stems from the resolution of the flow equations suitably linearized around a FP, determines the renormalization 
properties of the model at that FP. In fact, 
as discussed in \cite{Morris:1998da},
after establishing  the critical surface around the FP, spanned by the full set of
UV-attractive, or relevant,  directions, the continuum limit of the model 
is realized on the `renormalized trajectories' (RTs), i.e. RG flow trajectories, 
which land on the critical surface in the continuum limit $t \to\infty$.
The form of the effective action in terms of eigenfunctions around 
a fixed point  is
\small
\begin{equation}
S_k[\phi]=S_*[\phi]+\sum_{i=1}^n \alpha_i e^{-\theta_i t} \mathcal{O}_i[\phi]
\, ,\label{RT}
\end{equation}
\normalsize 
where  $S_*[\phi]$ is the FP action and the  operators 
$\mathcal{O}_i[\phi]$ are the eigenperturbations and  $\theta_i$ 
the corresponding eigenvalues. The sign of $\theta_i$ defines the nature 
of the eigenfunction, namely positive, negative or vanishing $\theta_i$
correspond respectively to relevant, irrelevant or marginal operators.
To explicitly determine the eigenvalues we express the variables 
$Y(x,t)$ and $z(x,t)$ as 
\small
\begin{align}
&Y(x,t)=Y^*(x)+\delta\, e^{-t\,\theta}h(x)\, ,\label{perturbY}\\
&z(x,t)=z^*(x)+\delta\, e^{-t\,\theta}f(x)\, .\label{perturbz}
\end{align}
\normalsize
where $h(x)$ and $f(x)$ represent small perturbations around the FP 
solution $Y^*$, $z^*$ (which are related to $u^*$, $w^*$ through 
Eqs. \eqref{numsol1}, \eqref{numsol2})
and  $\delta$ is a small number used to parametrize the expansion around 
the FP. Remarkably, as we already found out in the case of the FP, for a specific 
monomial form of $f(x)$ and $h(x)$, the linerarized flow equations become 
field independent and the problem is again reduced to a set of algebraic equations.

In fact, we consider  ${f(x)=s\,x^p}$ and ${h(x)= r\, x^q}$, 
with non-negative integers $p$ and $q$  and with 
constant $s$ and $r$.
Then, with 
these prescriptions, the linearized equations 
\eqref{eqfinY} and \eqref{eqfinz} 
(i.e. the coefficients of the  linear terms in $\delta$ when 
Eqs. \eqref{eqfinY} and  \eqref{eqfinz}
are expanded in powers of $\delta$,
after the insertion of Eqs. \eqref{perturbY}, \eqref{perturbz} ) become
\small
\begin{equation}
4-\frac{\eta  q}{2} -  e^{12 u^*}\,w^*\, \left(\frac{12\, s\, u^*\, 
x^{-q+p+4}}{r}+(q-1)\, q\right) =\theta \label{ddun}
\end{equation}
\normalsize
\small
\begin{align}
& 
\frac{16\, (\eta -1)\,(q-1)\, q\,r \,e^{12 u^*}\, u^*\, w^* 
\,(q+12 u^*-2)\, 
x^{q-p-4}}{(\eta -2) s} \nonumber\\
&+ \Bigg [
384 (\eta -2) (\eta -1)\, (u^* )^2\, (6 u^* +1)
+16 (\eta \, (3 \eta -7)+6)\, p\, u^*
\nonumber\\
&+(\eta -2)^2 p -(\eta -2)^2 p^2\Bigg]\;
\frac{e^{12 u^*}\, w^*}{(\eta -2)^2}- 2 -\frac{\eta}{2} 
\, (p+2) = \theta  \label{ddzn}
\end{align}
\normalsize
Only if ${q=p+4}$ these two equations become independent 
of $x$ and therefore, the spectrum  of eigenfunctions 
consists of the pairs 
${(f,h)=(s,r\,x^4)}$,  ${(f,h)=(s\,x^2,r\,x^6)}$, 
${(f,h)=(s\,x^4,r\,x^8)}$ 
and so on, where the coefficient $s$ and $r$ have to be
determined separately for each eigenfunction
from Eqs. (\ref{ddun}) and (\ref{ddzn}).

Then, for each value of $\eta$,  which selects a FP solution
$u^*(\eta), \, w^*(\eta)$ according to Fig. \ref{fig:line},
and for each eigenfunction ${(f,h)}$
we compute the parameters $s$ and $\theta$ from Eqs. 
\eqref{ddun} and \eqref{ddzn} 
(note that in these equations the parameter $r$ only appears 
in the ratio $(s/r)$ and therefore we can choose $r=1$ 
to normalize the eigenfunction ${(f,h)}$),
whose  non-linear structure provides 
either a pair of complex conjugate eigenvalues  
$\theta, \, \theta^*$, or a pair of real eigenvalues 
$\theta_1, \,\theta_2$.
Numerous eigenfunctions were tested, up to 
${(f,h)=(s\,x^{1000}, r\,x^{1004})}$, and in all cases 
the complex eigenvalues are found in the ranges 
$\eta<1$ and $2<\eta$, while the real eigenvalues 
in the interval  $1<\eta<2$.
These results are summarized in Figs. \ref{fig:mix1}, 
\ref{fig:mix2}.

In Fig. \ref{fig:mix1} the real and imaginary part of $\theta$
corresponding to $\eta<1$ (in blue), and to $2<\eta$ (in red), 
are plotted, for the three eigenfunctions corresponding to $q=4,6,8$
(we recall that the second solution for each value of $\eta$ is
$\theta^*$).

We observe that the eigenvalues associated with $q=4$ are 
different from all the others as they have negative real part
(which indicates irrelevant eigenfunctions) 
when $2<\eta$ and $0.96 \simeq \eta_c<\eta<1$,
and positive real part  (relevant eigenfunctions)
for $\eta<\eta_c$. 
Conversely, all other cases with $q>4$ show 
negative real part for $\eta<1$, and positive for $2<\eta$.

Remarkably, at ${\eta=0}$, our numerical 
determination of the eigenvalue associated with ${q=4}$ 
(the only one with positive real part), namely  
${\theta_{1,2}=2.919\pm i\, 3.923}$, 
coincides with the eigenvalue of the
asymptotically safe trajectory of the  Reuter FP  found 
in \cite{Bonanno:2012dg} for the $\mathrm{R}^4$ projection with 
$n=+\infty$.

Finally, in Fig. \ref{fig:mix2} the eigenvalues in the region 
$1<\eta<2$ are displayed. In this case, the two complex conjugate
solutions found in the other ranges of $\eta$, are replaced by 
two real solutions plotted respectively on the $x$ and $y$ axis
in the figure. Again, we observe a special role of the solutions
with $q=4$, because both real eigenvalues are positive for any 
value of $\eta$ in this range. 
All other solutions, corresponding to 
$q>4$, have only one positive eigenvalue in the 
restricted range $1<\eta<\overline\eta$ 
(with $\overline\eta \lesssim 2$), while the remaining 
solutions are negative.

Then, we emphasize the striking feature 
emerging from Figs. \ref{fig:mix1}-\ref{fig:mix2}, that at fixed 
$\eta$, all eigenfunctions share the same behavior, 
with the only exception of the one with $q=4$ which,
in  most cases with complex eigenvalues
shows opposite sign of the  real part of the eigenvalue  
with  respect to the other eigenfunctions,
while it shows at least one
eigenvalue with opposite sign, in the case of real eigenvalues.

\section{Conclusions}

The approximation scheme adopted in  this paper to analyze the
RG flow of a generalization of the conformally 
reduced Einstein-Hilbert action, yields a rich 
picture  of the UV critical manifold of the theory.

Specifically, in our approach the background field is treated 
in the so called single field approximation 
within the framework of the proper time flow, 
where the background  field is eventually {identified with the full field expectation.}
This scheme implies a distinction between the spectra of 
$\hat{\Box}$ and $\bar{\Box}$, and consequently it is possible 
to introduce the anomalous dimension $\eta$
through Eqs. (\ref{chipsi}) and (\ref{etaminustwo}), 
and $\eta$ is eventually determined by integrating  over 
the fluctuations of the conformal factor.  

Then, the generalization of the CREH action reported 
in Eq. (\ref{Sgen}), studied in this paper,
generates a  continuum of fixed points suitably parameterized 
by the anomalous dimension $\eta$ in Fig. \ref{fig:line}.
Each fixed point possesses a discrete spectrum of 
eigenoperators as 
shown in Figs. \ref{fig:mix1} and \ref{fig:mix2}. 

On one hand, these findings represent the improvement produced 
by our approximation scheme with respect to the analysis of 
the plain CREH truncation  \cite{Bonanno:2012dg} 
where a single FP, namely
the Reuter FP related with the property of asymptotic safety,
was spotted. 
On the other hand, the UV critical manifold illustrated here 
presents an evident dissimilarity with respect to the 
one derived  in \cite{Dietz:2016gzg} where 
a full background  independent approximation scheme 
(realized by resorting to modified split Ward identities)
was adopted and a continuum of fixed points, supporting both 
a discrete and a continuous eigenoperator  spectrum is found. 
The partial incongruity of the two results, 
which is certainly to be addressed to the different 
approaches adopted in the two cases, in our opinion 
is an clear indicator of the sensitivity of the predictions 
to the particular procedure and approximation scheme selected.

Turning to the results of the analysis discussed above,
if we attempt to establish the 
physical significance of our solutions
from the eigenfunction analysis, rather than 
from the determination of the sign of the FP shown in
Fig. \ref{fig:line}, we should discard FPs and 
eigenfunctions corresponding to 
$\eta>2$, as in this case we find an infinite number of 
relevant directions (those with $q>4$) and just an 
irrelevant one ($q=4$) and this would correspond to 
a not predictive theory.

Conversely, the line of FPs in  the range  
$ \eta < \eta_c \approx 0.96$,
has the desired  properties of yielding both  positive 
Newton's Constant and Cosmological Constant,
with complex eigenvalues associated to the eigenfunction 
solutions of the  linearized flow equation  around each FP,
but with just one relevant solution (i.e. with positive 
real part of the eigenvalue) 
that produces the renowned spiral behavior.
In particular, the FP at $\eta=0$ correctly reproduces 
the main features  of the Reuter FP.
Moreover, it is worth noticing that the presence 
of an endpoint in a continuous line of FPs,
analogous to $\eta_c$ in our analysis,
in the case of the two-dimensional 
Kosterlitz-Thouless transition is directly related
to the  universal, physically measurable, property 
of the spin-stiffness jump \cite{Kosterlitz_1972,Nelson1977}.

Finally, the range $1<\eta<2$  has infinite relevant 
directions (and therefore it corresponds again to 
a not predictive theory that must be discarded),  
except for the small range $\overline \eta< 
\eta < 2$ where only the two real eigenvalues with 
$q=4$ are positive. 
Therefore, this small range of $\eta$ produces a 
peculiar structure, different from the spiral behavior 
so far observed, which certainly deserves a more accurate
investigation, in order to clarify whether it corresponds to
a genuine physical effect and not to an artifact of the 
adopted approximation.
 

\section*{Acknowledgements}
\noindent
MC thanks the National Institute of Astrophysics (INAF) 
section of Catania together with the INAF - Catania Astrophysical 
Observatory for their hospitality during the preparation of the 
manuscript. MC acknowledges support from National Institute of 
Nuclear Physics (INFN). MC also wishes to thank Sergio Cacciatori 
for his invaluable help with the 
`phiboxing'.
This work has been carried out within the INFN project FLAG.
\bibliographystyle{unsrt}

\end{document}